\input{aipcheck}

\documentclass[
    ,final            
  ]
  {aipproc}
\layoutstyle{6x9}
\usepackage{graphicx,epsfig,rotating}        

\begin{document}

\title{High-p$_{T}$ azimuthal correlations of neutral strange baryons and mesons in STAR at RHIC}

\classification{25.75.-q, 25.75.Gz}
\keywords      {identified correlations, high-$p_T$ processes, neutral strange baryons and mesons}

\author{Jana Bielcikova for the STAR Collaboration}{
address={Physics Department, Yale University,  P.O. Box 208120, New Haven, CT 06520-8120, USA}}

\begin{abstract}
We present results on two-particle azimuthal  correlations of high-$p_T$ neutral strange baryons 
($\Lambda$,$\bar{\Lambda}$) and mesons ($K^0_S$) associated with non-identified 
charged particles in d+Au and Au+Au collisions at $\sqrt{s_{NN}}$~=~200~GeV. In particular, 
we discuss properties of the  near-side yield of associated charged particles as a function of centrality,
transverse momentum and $z_T$, as well as possible baryon/meson and particle/antiparticle 
differences. The results are compared to the proton and pion 
triggered correlations and to fragmentation and recombination models.
\end{abstract}

\maketitle


\section{Introduction}

Partonic energy loss is predicted to be a sensitive probe of the matter created in high energy heavy-ion 
collisions because its magnitude depends strongly  on the color charge density of the medium traversed. 
Partons originating from hard scattering of quarks or gluons from the colliding nuclei fragment into jets.  
As a direct measurement of jets in nuclear collisions is difficult  due to the high multiplicities 
of emitted particles, azimuthal correlations of particles with large transverse momentum ($p_T$) 
are commonly used to study jet related processes. Identified two-particle correlations are expected 
to provide additional information on jet quenching, as well as the baryon-meson puzzle 
and particle production mechanisms at RHIC energies.  In this paper, we discuss properties 
of the near-side associated yield of charged particles for identified strange baryon 
($\Lambda$, $\bar{\Lambda}$) and meson (K$^0_S$) trigger particles. 

\section{Data analysis}
The analysis presented in this paper is based on d+Au and Au+Au collisions 
at $\sqrt{s_{NN}}$~=~200~GeV, measured by the STAR experiment. 
STAR is a multi-purpose spectrometer comprised of several detectors inside a large 
solenoidal magnet with a magnetic field of 0.5~T. In this analysis, we use both charged 
particle tracks measured by the Time Projection Chamber (TPC) with full azimuthal coverage, 
together with $\Lambda$, $\bar{\Lambda}$ and K$^0_S$ particles reconstructed 
by a topological analysis from their decay products. 

In the azimuthal correlations presented here, we distinguish several trigger particle species (charged particles, 
$\Lambda$, $\bar{\Lambda}$, and K$^0_S$) while the associated particles are 
always charged particles with $p_T$~=~(1--2)~GeV/$c$.
The measured azimuthal distributions are normalized to the number of trigger particles 
and corrected for the reconstruction efficiency of associated charged particles
which, in the $p_T$ range studied, varies from 70-84\% depending on centrality. 
The data are fit with two Gaussians on top of a flat background in d+Au 
and elliptic flow modulated background in Au+Au collisions, respectively. 
The yield of associated particles is calculated as the area under the Gaussian peak.
The uncertainties in the elliptic flow subtraction result in about a 30$\%$ 
systematic error on the extracted associated yield.

\begin{figure}[t!]
\begin{tabular}{lr}
  \includegraphics[height=5.9cm]{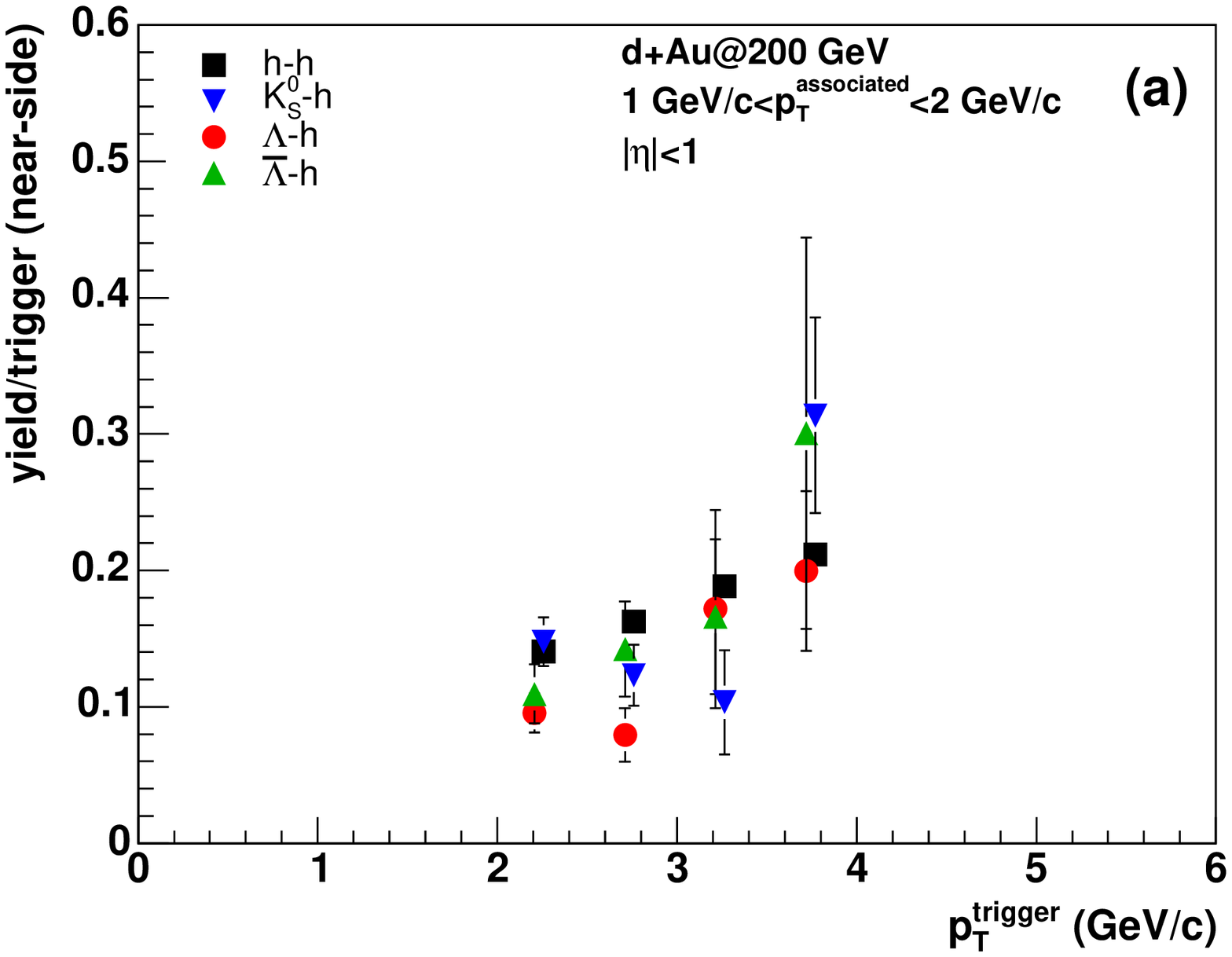}
&
\hspace{-0.6cm}
 \includegraphics[height=5.9cm]{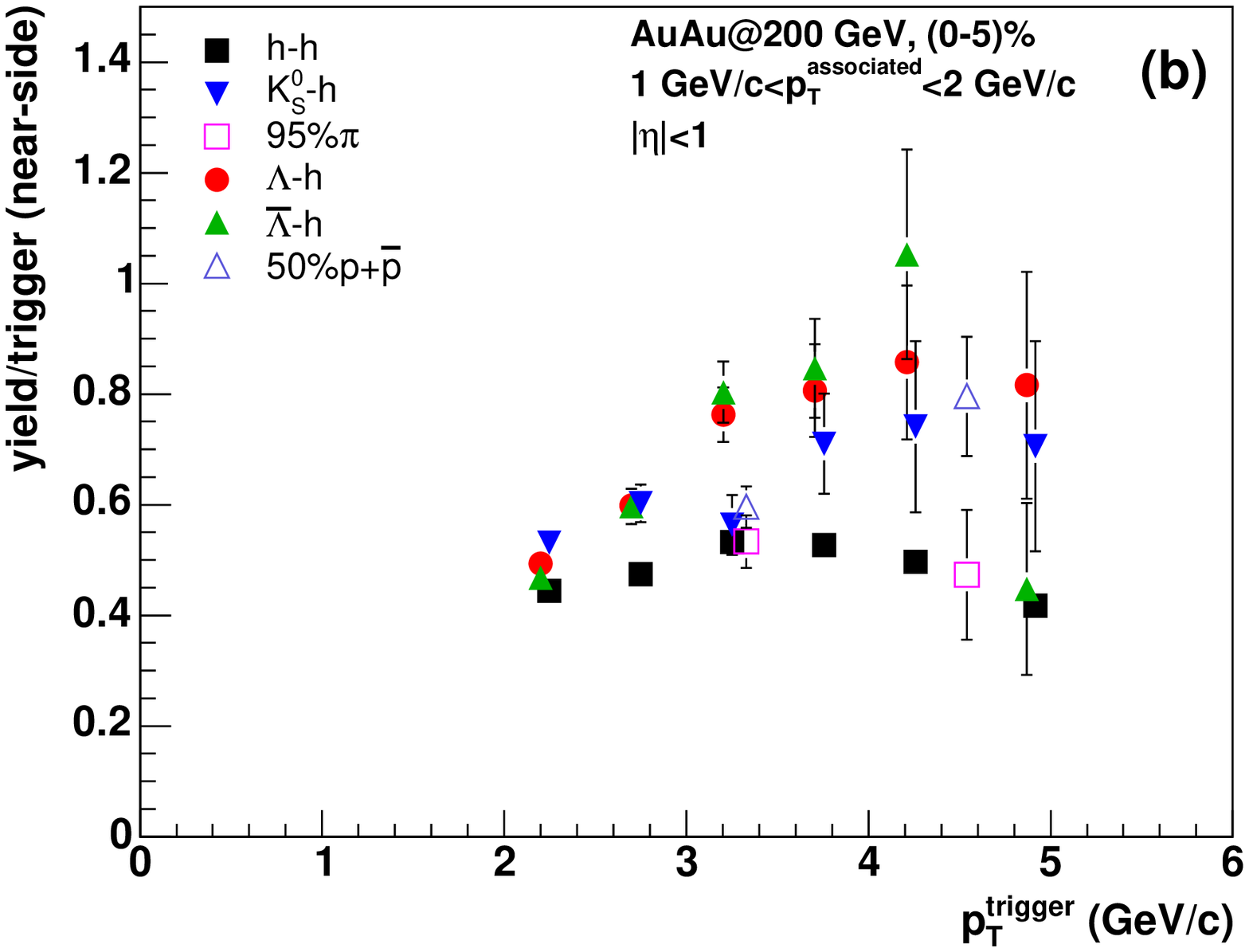}
\vspace{-0.4cm}
\end{tabular}
\caption{Near-side associated charged particle yield as a function of the transverse momentum of the trigger particle, $p_T^{trigger}$, in d+Au (a) and central Au+Au collisions (b) at $\sqrt{s_{NN}}$~=~200~GeV. 
The data points for baryons are offseted by 50~MeV/$c$ for a better view.}
  \label{pttrig}
\end{figure}

\section{Results}
Fig.~\ref{pttrig} shows the near-side associated yield of charged particles with $p_T$~=~(1--2)~GeV/$c$ 
as a function of the transverse momentum of the trigger particle, $p_T^{trigger}$. 
The near-side yield in central Au+Au collisions is about 
3-4 times larger than in d+Au collisions, independent of trigger species and $p_T^{trigger}$.
To look for possible baryon/meson differences, we have included in Fig.~\ref{pttrig} the results
for $\pi^+$+$\pi^-$ (95\% purity) and $p$+$\bar{p}$ (50\% purity) trigger particles which are 
identified by the relativistic rise 
of ionization energy loss in the TPC~\cite{Ulery:2005cc}. Although the near-side yield does not show any 
significant difference between trigger particle species in either collision system, 
there is a hint of a trend of a baryon/meson splitting in Au+Au collisions which is not apparent in the d+Au data.

\begin{figure}[t!]
\begin{tabular}{lr}
 \includegraphics[height=5.9cm]{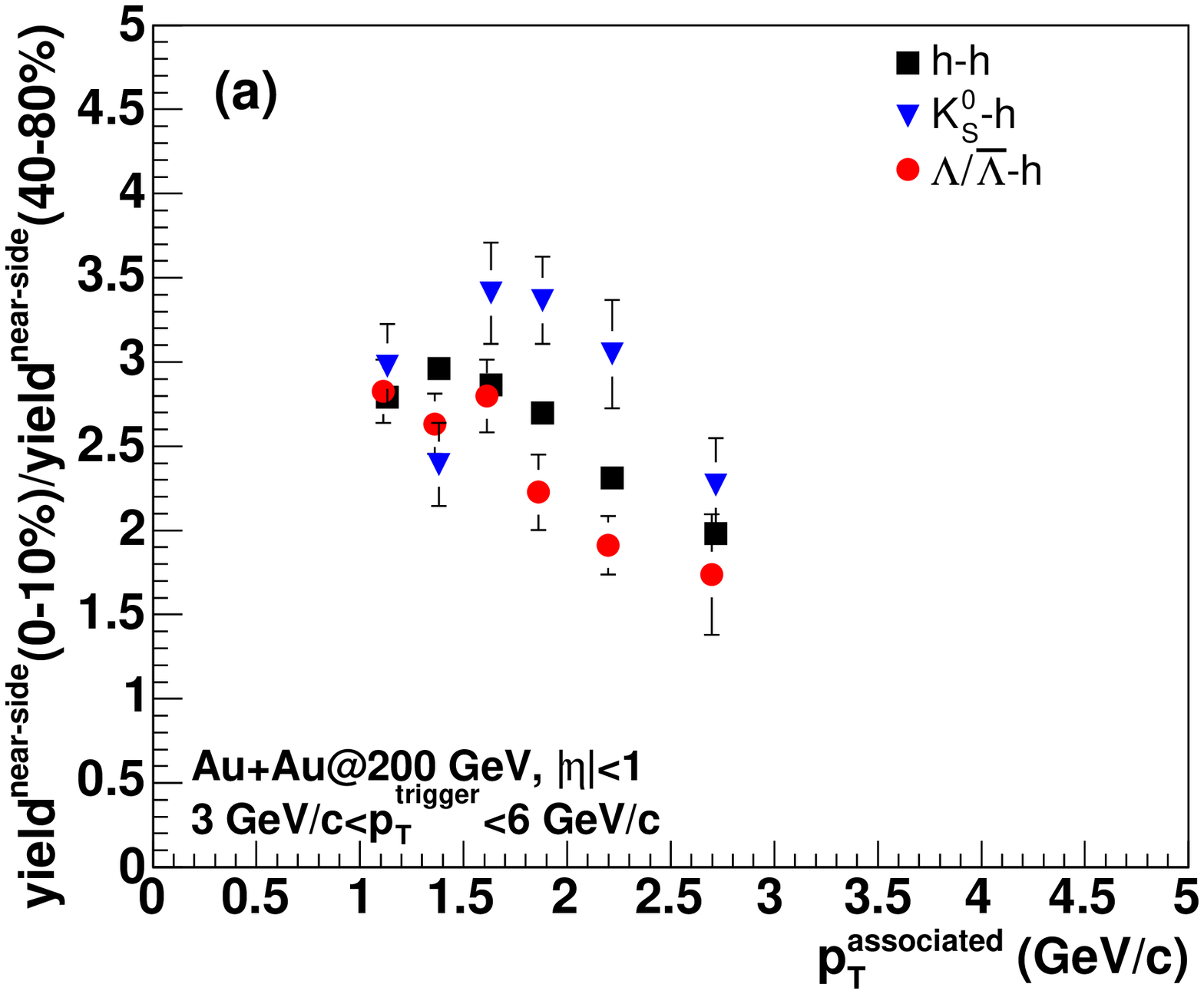}
&
\includegraphics[height=5.9cm]{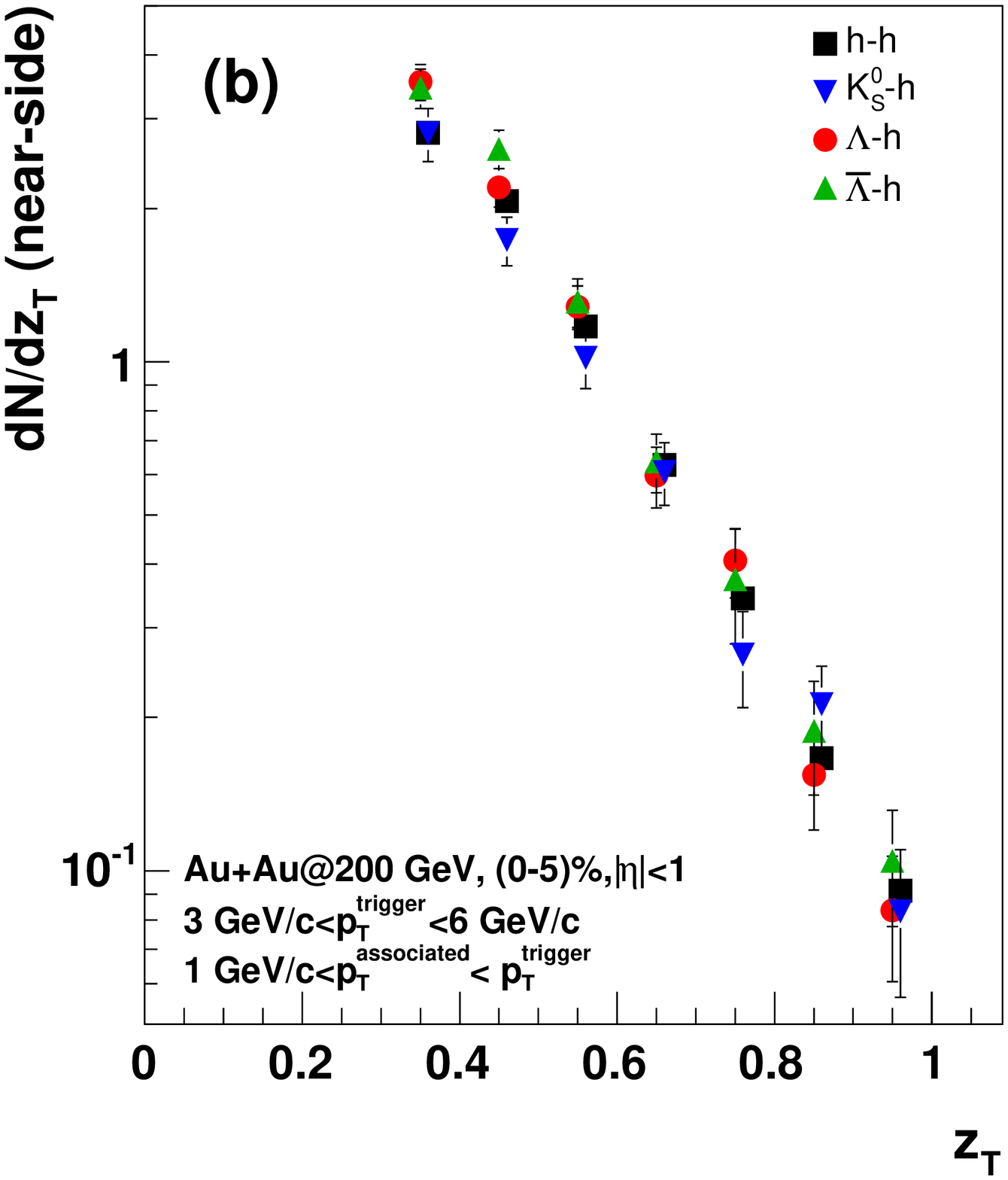}

\vspace{-0.4cm}
\caption{The ratio of near-side associated charged particle yield in central (0-10$\%$) 
and peripheral (40-80$\%$) Au+Au collisions as a function of $p_T^{associated}$ (a). 
$z_T$ dependence of near-side charged particle associated yield in central Au+Au collisions  
at $\sqrt{s_{NN}}$~=~200~GeV. 
The data points for baryons are offseted by 50~MeV/$c$ for a better view (b).}
\label{rcp-hwa-ztdep}
\end{tabular}
\end{figure}

Next, we discuss the centrality dependence of the near-side associated yield. In order to make a comparison 
with a parton recombination model~\cite{Hwa:2005ui}, we have calculated the ratio of the near-side yield 
in central (0-10\%)  to peripheral (40-80\%) Au+Au collisions for $p_T^{trigger}$~=~3--6~GeV/$c$. 
The ratio decreases from about 3 at $p_T^{associated}$~=~1~GeV/$c$ 
to 2 at $p_T^{associated}$~=~3~GeV/$c$ (cf.~Fig.~\ref{rcp-hwa-ztdep}(a)).
This behavior, as well as the large magnitude of the ratio, is in line with the 
model expectations and points toward a significant role of thermal-shower recombination in Au+Au 
collisions~\cite{Hwa:2005ui} .
However, as long range pseudo-rapidity correlations play a significant role in Au+Au collisions 
for $p_T<$~3~GeV/$c$~\cite{putschke} and they are not included in the model, this agreement is only qualitative. 
Again, there is a hint of a trend of a baryon/meson splitting although the errors are quite large.

The near-side associated yield can be connected to fragmentation. It was proposed
to study charged hadron-triggered fragmenation functions expressed in terms 
of the variable $z_T~=~p_T^{associated}/p_T^{trigger}$~\cite{Wang:2003mm}, an approximation of the commonly 
used variable $z~=~p_T/p_T^{parton}$. Here, we go one step further and present identified hadron-triggered 
fragmentation functions. Fig.~\ref{rcp-hwa-ztdep}(b) shows the fragmentation functions 
for $p_T^{trigger}=$~3~--~6~GeV/$c$ in central Au+Au collisions. In the $p_T$ range studied,  
the spectral shape is similar for various trigger particle species within the statistical errors.

\section{Conclusions}
We have reported results on near-side associated yield of charged particles for charged hadron and 
neutral strange baryon/meson trigger particles.
There is a large increase of the near-side yield from d+Au towards central Au+Au collisions where thermal-shower recombination dominates. 
However, in the $p_T$ range studied we do not observe any appreciable baryon/meson 
or particle/antiparticle differences in either collision system, although there 
is a hint of a trend of baryon/meson difference 
in Au+Au collisions. Ongoing studies with larger statistics will allow access to a higher $p_T$ 
range which, in combination with a study of long range pseudo-rapidity correlations for strange 
trigger particles, will bring new insights into this problem.    




\bibliographystyle{aipproc}   

\end{document}